\documentclass[twocolumn,superscriptaddress,preprintnumbers,amsmath,amssymb]{revtex4}
\usepackage{amsmath,graphicx,enumerate,subfigure}
\newcommand{\eqnref}[1]{Eq.~(\ref{#1})}
\newcommand{\figref}[1]{Fig.~\ref{#1}}

\usepackage{dcolumn}% Align table columns on decimal point
\usepackage{bm}% bold math
\begin{document}

\preprint{APS/123-QED}

\title{Bound states near a moving charge in a quantum plasma}% Force line breaks with \\

\author{D. Else}
\affiliation{School of Physics,
The University of Sydney, New South Wales 2006, Australia}

\author{R. Kompaneets}
\affiliation{School of Physics,
The University of Sydney, New South Wales 2006, Australia}

\author{S. V. Vladimirov}
\affiliation{School of Physics, The University of Sydney, New South Wales 2006, Australia}
\affiliation{Joint Institute for High Temperatures, Russian Academy of Sciences, 
ul. Izhorskaya 13 bld 2, 127412 Moscow, Russia}

\date{\today}% It is always \today, today,
             %  but any date may be explicitly specified

\begin{abstract}
{It is investigated how the shielding of a moving point charge in a one-component fully degenerate fermion plasma
affects the bound states near the charge at velocities smaller than the Fermi one. The shielding is accounted for by using the Lindhard dielectric function,
and the resulting potential is substituted into the Schr\"odinger equation in order to obtain the energy levels. 
Their number and values are 
shown to be primarily determined by the value of the charge and the quantum plasma coupling parameter, while the
main effect of the motion is to split certain energy levels.
This provides a link between quantum plasma theory and possible measurements
of spectra of ions passing through solids.}
\end{abstract}

\maketitle

\section{Introduction}

Quantum plasmas have recently received a rapidly growing interest \cite{manfredi-fields.inst.commun-2005, melrose-book-2008, fortov-book-2008,
shukla-nature-2009, shukla-eli-plasma.phys.control.fusion-2010, shukla-eli-phys.usp-2010, manfredi-her-yin-book-2010}. 
The motivation is related, in particular, to nanostructured metallic and semiconductor 
materials \cite{manfredi-her-yin-book-2010, manfredi-her-appl.phys.lett-2007}, laser-plasma systems \cite{fortov-khrapak-book-2005, marklund-shu-rev.mod.phys-2006}, 
astrophysics \cite{brodin-mar-eli-phys.rev.lett-2007}, and ultracold plasmas \cite{li-tan-gal-phys.rev.lett-2005}. 
Quantum plasmas can be described by various models including the hydrodynamic one with the Bohm potential \cite{manfredi-haa-phys.rev.b-2001,
manfredi-her-yin-book-2010}, kinetic Wigner-Poisson model (a quantum analogue of the classical Vlasov-Poisson system) \cite{manfredi-haa-phys.rev.b-2001,
manfredi-her-yin-book-2010}, 
quantum relativistic plasmadynamics \cite{melrose-book-2008}, and other models \cite{manfredi-her-yin-book-2010, kremp-sch-krae-book-2005}. 
Based on these and similar approaches, recently there have been studies of quantum effects on various plasma phenomena 
including two-stream instability \cite{haas-bre-shu-phys.rev.e-2009}, ion-acoustic waves \cite{ali-mos-shu-phys.plasmas-2007, mushtaq-mel-phys.plasmas-2009}, 
spin effects on the plasma dispersion \cite{marklund-bro-phys.rev.lett-2007, brodin-mar-zam-phys.rev.lett-2008}, whistlers \cite{misra-bro-mar-phys.rev.e-2010}, 
Bernstein modes \cite{eliasson-shu-phys.plasmas-2008}, Zakharov dynamics \cite{marklund-phys.plasmas-2005}, and thermodynamic 
properties \cite{filinov-bon-for-j.phys.a.math.gen-2006, filinov-lev-bot-j.phys.a.math.theor-2009}.

As a part of research in this field, there have been numerous works to link theoretical models to practice. 
For instance, already in 1956, Watanabe showed that his measurements of energy losses and
scattering angles of electrons passing through a thin metallic film are in a remarkable agreement with the quantum plasma 
dispersion relation that can be derived from the Wigner-Poisson model \cite{watanabe-j.phys.soc.japan-1956}. As regards 
recent works, Manfredi and Hervieux performed quantum plasma simulations that reproduce low frequency nonlinear 
oscillations revealed in transient reflection experiments on thin films \cite{manfredi-her-phys.rev.b-2005}. Furthermore, 
Marklund et al. demonstrated that the quantum collisionless damping of surface plasmons poses a fundamental size limit for 
plasmonic devices \cite{marklund-bro-ste-europhys.lett-2008}. 

Another link between theory and practice is related to spectra of ions moving in quantum plasmas. 
Indeed, a plasma shields the Coulomb potential of a moving charge and thus affects the bound electron states near it. 
This was demonstrated experimentally in the regime where the charge velocity exceeds the Fermi one \cite{bell-bet-pan-j.phys.b-1976, jakubassa-j.phys.c-1977}. 
In this regime, the effect has been 
studied in a number of theoretical papers \cite{jakubassa-j.phys.c-1977, neelavathi-rit-bra-phys.rev.lett-1974, day-phys.rev.b-1975, 
echenique-rit-phys.rev.b-1980, mazarro-ech-rit-phys.rev.b-1983, echenique-bra-rit-phys.rev.b-1986}.

The object of this Letter is to investigate this effect for charge velocities {\it smaller} than the Fermi one. 
In this regime, as opposed to that where the charge velocity exceeds the Fermi one, two factors may be particularly significant. 
The first one is the spread of velocities in a fully degenerate distribution, as the plasma can no longer be considered as cold or 
almost cold. The second one is the quantum recoil/tunneling. Its role in the shielding of a moving charge was studied in detail in Ref.~\cite{else-kom-vla-phys.rev.e-2010}.

\section{Model}
We consider a point charge $+Ze$ representing an ion and
moving through a plasma at constant velocity $\mathbf{v}$.
Here, $e$ is the elementary charge. The plasma is assumed to be a fully degenerate electron gas
immersed in a neutralizing background.
We adopt the frame of the moving ion and use it as our origin of coordinates.

In describing the quantum-mechanical dynamics of a single electron near the ion in the presence of the plasma,
we consider the response of the plasma to be solely due the ion; this is justified when the plasma responds much more strongly to the ion than the electron, as will be the case \cite{echenique-bra-rit-phys.rev.b-1986} for $Z \gg 1$. Thus, the problem is reduced to solving the time-independent Schr\"odinger equation for the one-electron wavefunction $\psi(\mathbf{r})$, namely
\begin{equation}
\label{schrodinger_eqn}
-\frac{\hbar^2}{2m_e} \nabla^2 \psi - e \varphi(\mathbf{r}) \psi = E\psi,
\end{equation}
where $m_e$ is the mass of the electron and $\varphi(\mathbf{r})$ is the electrostatic potential generated by the ion in the plasma. 
Note that, due to the rotational symmetry of the potential about the axis of
motion, we can impose the dependence on the azimuthal angle $\phi$ in spherical
or cylindrical coordinates to be of the form $\psi \propto e^{im\phi}$ for some
integer value of the magnetic quantum number $m$. The binding energies will be identical for positive and negative $m$.

The potential $\varphi(\mathbf{r})$ can be calculated in the linear approximation by:
\begin{multline}
\label{linear_potential}
\varphi(\mathbf{r}) = \frac{Ze}{4\pi\epsilon_0}\left(\frac{1}{r}  \right.
\\
\left. + \frac{1}{2\pi^2}\int
\frac{\exp(i\mathbf{k}\cdot\mathbf{r})}{k^2}\left[\frac{1}{D(\mathbf{k}\cdot\mathbf{v}, \mathbf{k})} - 1\right]d\mathbf{k} \right),
\end{multline}
where $D(\omega,\mathbf{k})$ is the dielectric response function of the plasma and $\epsilon_0$ is the permittivity of free
space. Here we use Lindhard's dielectric function 
\cite{lindhard-1954, lifshitz-book}
for the response of fully degenerate electrons, which reads
\begin{subequations}
\label{dielectric_quantum}
\begin{multline}
D(\omega, \mathbf{k}) = 1 + \\
\frac{3\omega_p^2}{k^2v_F^2}\frac{F(\omega +
i\nu + a, k) - F(\omega + i\nu - a, k)}{2a},
\end{multline}
where
\begin{gather}
a = \frac{\hbar k^2}{2m_e}, \\
F(\Omega, k) = \frac{\Omega}{2} + \frac{(kv_F)^2 - \Omega^2}{4kv_F} 
\ln\left(\frac{\Omega + kv_F}{\Omega - kv_F}\right),
\end{gather}
\end{subequations}
$v_F = (3\pi^2)^{1/3} \hbar n_e^{1/3}/m_e$ is the electron Fermi velocity, 
$\omega_p = \sqrt{n_e e^2/(m_e \epsilon_0)}$ is the electron plasma frequency, and $n_e$ is the electron number density; $\nu$ represents an infinitesimal positive number. 
The Lindhard response function can be derived from the Wigner-Poisson system and includes both the quantum tunneling and the degeneracy.
The potential resulting from the Lindhard dielectric function has been studied in Refs.~\cite{day-phys.rev.b-1975,mazarro-ech-rit-phys.rev.b-1983,else-kom-vla-phys.rev.e-2010}.
The applicability of our model is analyzed in the discussion section.

We have solved \eqnref{schrodinger_eqn} numerically by finite-difference discretization,
with the resulting sparse matrix eigenvalue problem solved using the \textsc{arpack} library \cite{arpack_users_guide}. The potential $\varphi(\mathbf{r})$ is found by numerical integration of \eqnref{linear_potential}.
The bound states in this problem are determined (up to scaling) by the following dimensionless parameters: 
$Z$, the atomic number of the ion; the ratio $v/v_F$; and $H =
\hbar\omega_p/(m_e v_F^2)$, which characterizes the strength of the quantum recoil
and is also related to the plasma coupling parameter.

\section{Results}
Note that as $H \to 0$, the quantum tunneling becomes unimportant to the screening so that the non-dimensionalized potential [i.e.\ with position
normalized to $\lambda = v_F/\omega_p$, and the
potential normalized to $Ze/(4\pi\epsilon_0 \lambda)$] tends towards the ``semiclassical'' form.
The ``semiclassical'' form implies that it can be obtained from the classical Vlasov-Poisson system
with the fully degenerate velocity distribution. In this limit, the potential
surrounding a \emph{stationary} test charge 
tends to the Debye-H\"uckel one,
with the resulting screening length being equal to the Thomas-Fermi length $\lambda/\sqrt{3}$.

Once the non-dimensionalized potential has been specified, the bound states depend (up to scaling) only on $ZH$.
The physical meaning of the product $ZH$ is that it is equal (up to a constant coefficient of the order of unity)
to $\lambda/(a_0/Z)$, where $a_0 = 4\pi\epsilon_0 \hbar^2/(m_e e^2)$ is the Bohr radius. Note that 
$a_0/Z$ is the characteristic scale of the ground state of the ion in vacuum.
Therefore it is informative to consider the limit $H \to 0$, $ZH$ held constant.

\begin{figure}
\includegraphics[width=8.4cm]{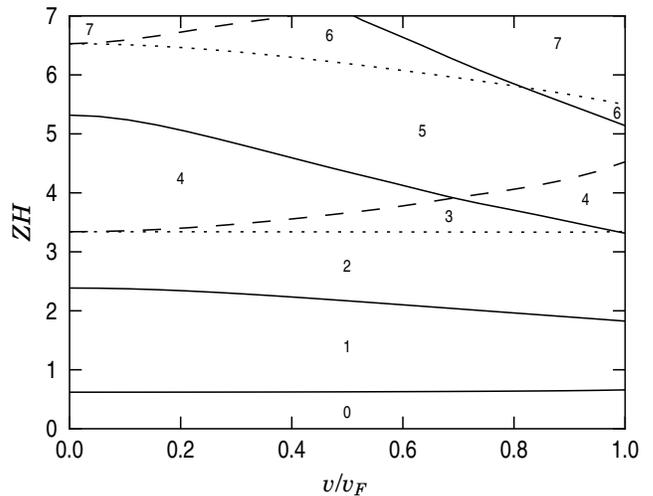}
\caption{\label{plot_state_boundaries_semiclassical}The number of distinct
energy levels (shown by numerals 0, 1, $\ldots$, 7) as a function of $v/v_F$ (horizontal axis) and $ZH$ (vertical axis), 
in the limit $H \to 0$, $ZH$
held constant. The lines indicate
the boundaries between the areas corresponding to different numbers of distinct energy levels. The line style reflects the quantum numbers (with the $l$
quantum number determined by tracing back to $v = 0$) of the bound state that appears when the line is crossed: solid for $l=0,m=0$; dashed for $l=1,m=0$; dotted for $l=1,m=\pm 1$.}
\end{figure}

\begin{figure}
%\subfigure[The ratio of the potential $\varphi(\mathbf{r})$ to the unscreened Coulomb potential $Ze/(4\pi\epsilon_0 r)$. The contours are equally spaced, and the thicker contour corresponds to the boundary between positive and negative potential.]
%\includegraphics{plot_potential_v_0_9_contour.eps}
%\subfigure[The wavefunction $\varphi(\mathbf{r})$ for one bound state, which in the static case would correspond to the quantum numbers $n=3,l=m=0$.]
\includegraphics[width=8.4cm]{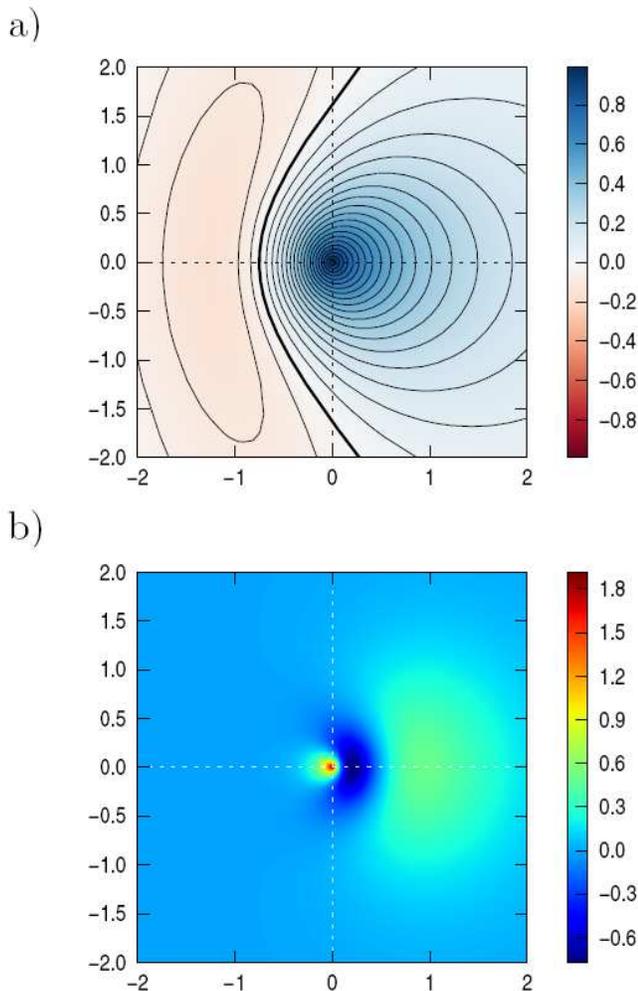}
\caption{The potential generated by the ion, at $v = 0.9v_F$ in the limit $H \to
0$; and one of the bound states in this potential, with $ZH$ held constant and
equal to 6.
The ion is located at the origin and moving to the right; the coordinates are
in units of $\lambda = v_F/\omega_p$.
(a) The ratio of the potential $\varphi(\mathbf{r})$ to the unscreened Coulomb potential $Ze/(4\pi\epsilon_0 r)$. 
The contours are equally spaced, and the thicker contour corresponds to the boundary between positive and negative potential. (b)
The wavefunction $\psi(\mathbf{r})$ for a particular bound state, which can
be continuously traced back to the bound state near a stationary test
charge with quantum
numbers $n=3,l=m=0$.
\label{plot_potential_and_wavefunction}}
\end{figure}

\figref{plot_state_boundaries_semiclassical} shows the number of energy levels in the limit $H \to 0$, $ZH$ held constant. 
As $ZH$ increases, the ratio of the screening length to the dimensions of individual bound states becomes larger, and 
thus more bound states appear and the binding energy of levels increases, tending to the bare Coulomb value $|E| = 0.5Z/n^2\,\mathrm{a.u.}$ 
(where $n$ is the principal quantum number) as $ZH \to \infty$. The splitting of energy levels occurs because the isotropy of the potential
is broken.
\figref{plot_potential_and_wavefunction} 
shows how a state which at $v=0$ is isotropic is distorted by the changes to the potential.

In general, the effect of the motion on the potential is greater screening (and the development of a repulsive potential)
behind the moving ion (and to a lesser extent in the direction perpendicular to the motion), and decreased screening in front of the ion \cite{else-kom-vla-phys.rev.e-2010}. 
Thus, since states with different angular quantum numbers (i.e. $l$ and $m$) have different angular dependencies, they are affected by the anisotropy of the potential in 
different ways, as can be seen in \figref{plot_state_boundaries_semiclassical}. In particular, some of the boundary curves in \figref{plot_state_boundaries_semiclassical} 
cross, with the result that for particular values of $ZH$, there can be narrow ranges of $v/v_F$ in which a bound energy level disappears, 
or an additional level appears. However, as states become more tightly bound, their spatial extent becomes much smaller than $\lambda$, and they cease to be 
greatly affected by the motion.

Interestingly, the minimal value of $ZH$ at which at least one bound state exists is almost independent on $v/v_F$
and is about $0.6$. At the same time, the value of $ZH$ at which the second level appears shows a more noticeable dependence on velocity.
This value decreases from about 2.3 at zero velocity to about 1.8 at the Fermi velocity and thus changes by about 20\%.

\begin{figure}
\includegraphics[width=8.4cm]{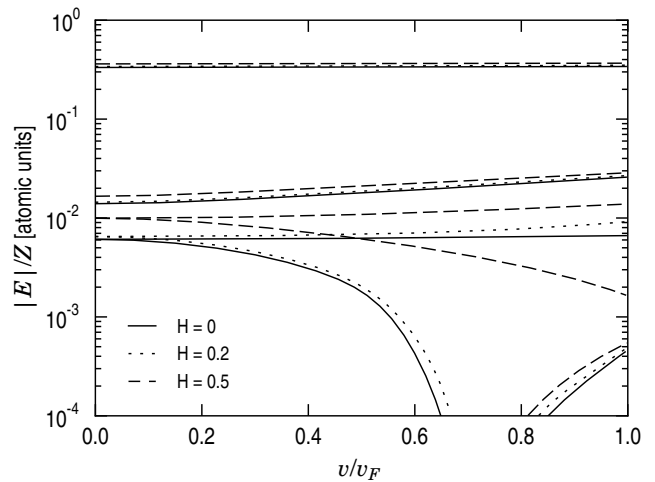}
\caption{\label{plot_binding_energies}The binding energies (divided by the
atomic number $Z$) as a function of
velocity for different values of $H$, but all with $ZH \approx 3.82$.
Different lines of the same style correspond to different bound states for the same parameters; all possible bound states
are shown for given parameters.}
\end{figure}

In the fully quantum case of finite $H$ (but less than unity), the quantum tunneling does not introduce qualitative changes to the above, though
it can make noticeable quantitative changes. This can be seen in \figref{plot_binding_energies}. It shows that
for a fixed $ZH$ the binding energies normalized by $Z$ increase with $H$ and tend to their semiclassical values at $H \to 0$.
Thus, the main effect of the quantum tunneling on \figref{plot_state_boundaries_semiclassical} would be to displace the boundary curves down.

\section{Discussion}
The use of our model is limited because 
it does not include correlations of plasma electrons, nonlinear and relativistic plasma effects as well as the plasma response to the bound electron. Let us discuss the 
corresponding applicability limits in terms of the parameters $H$ and $Z$.

The neglect of correlations of plasma electrons is justified 
when $H \ll 1$. Indeed, the use of the collisionless mean-field approximation requires that
the quantum plasma coupling parameter is small. The latter is defined as
$\Gamma=[e^2n^{1/3}/(4\pi \epsilon_0)]/[m_e v_F^2/2]$ which is the ratio of the characteristic potential energy of interaction between 
neighboring electrons to the Fermi energy. The parameter $\Gamma$ can be expressed as
$\Gamma = H^2 (3 \pi^2)^{2/3}/(2\pi)$, so we get the requirement $H \ll 1$.
In the limit $H \to 0$, however, the quantum tunneling disappears \cite{manfredi-haa-phys.rev.b-2001}. Thus, the fact that our model
includes the quantum tunneling but does not include particle
correlations makes it inconsistent in a certain sense. Nevertheless,
it is widely used in the literature, i.e. at small $H$ the
quantum tunneling is believed to be more important than particle
correlations.

The use of the linear response formalism [Eq.~(\ref{linear_potential})] requires $H \ll 1/Z^{1/3}$.
This limitation can be derived from the condition that the 
potential energy of the interaction of a plasma electron with the test charge at the length 
$\lambda$ is much smaller than the Fermi energy.

The relativistic plasma effects are negligible when $H \gg 0.06$. This condition follows from the requirement that 
the Fermi velocity is much smaller than the speed of light.
The latter requirement can be written as $H \gg 2 \sqrt{\alpha/(3\pi)} \approx 0.06$, where $\alpha = e^2/(4\pi\epsilon_0 \hbar c) \approx 1/137$ is the fine structure constant.

The plasma response to the bound electron should be unimportant when $Z \gg 1$. In this case the ion 
should generate a significantly stronger plasma response due to its larger charge.
The requirement $Z \gg 1$ is also evident from the results of Ref.~\cite{echenique-bra-rit-phys.rev.b-1986}.

The four above requirements can be easily satisfied at small $H$ and large $Z$. 
Note that the relativistic restriction
is formally not met in the limit $H \to 0$, $ZH$ held constant.
This means that taking this limit within our model gives results that apply at not extremely small,
though sufficiently small, values of $H$.

It follows that metals are at the edge of applicability of our model. For instance, for aluminum the parameter $H$ is about 0.7,
as can be deduced from its Fermi energy, 11.7 eV. This value of $H$ suggests that
nonideal plasma effects \cite{fortov-khrapak-book-2005, fortov-book-2008} can provide a non-negligible contribution. 
Nevertheless, the Lindhard dielectric function can apply quite well in metals, 
as evidenced by the experiment of Watanabe \cite{watanabe-j.phys.soc.japan-1956}. 

\section{Conclusion}
We have studied how the shielding a moving charge in a fully degenerate electron gas
affects the bound electron states near the charge for velocities smaller than the Fermi velocity. 
We accounted for the shielding by using the Lindhard dielectric function.

The main result is that the number of distinct energy levels is primarily determined by the parameter $ZH$, 
while the main effect of the motion is to split certain energy levels. 
In the semiclassical limit, i.e. $H \to 0$, $ZH$ held constant,
the 
minimal value of $ZH$ at which at least one bound state exists is almost independent of velocity and is about 0.6,
while the value of $ZH$ at which the second level appears varies from 2.3 to 1.8 with velocity.
In the fully quantum case of finite $H$ (but less than unity),
the quantum tunneling does not introduce qualitative changes, but it can introduce 
noticeable quantitative changes. They are that the binding energies normalized by $Z$ increase with $H$,
for a fixed $ZH$.

This provides a link between quantum plasma theory and possible measurements of 
spectra of ions passing trough solids. For instance, the measured positions of spectral lines
as well as their splitting could be compared to theoretical values. 
This might allow, in particular, assessing the role of the effects not included
in the linear mean-field model, for various conditions.

\begin{acknowledgments}
The work was partially supported by the
Australian Research Council.
R.K. acknowledges the receipt of a Professor Harry Messel Research Fellowship
funded by the Science Foundation for Physics within the University
of Sydney.
\end{acknowledgments}

\end{document}